\documentclass[letterpaper]{article} 
\usepackage{aaai23}  
\usepackage{times}  
\usepackage{helvet}  
\usepackage{courier}  
\usepackage[hyphens]{url}  
\usepackage{graphicx} 
\usepackage{natbib}  
\usepackage{caption} 
\usepackage{algorithm}
\usepackage{algorithmic}

\usepackage{newfloat}
\usepackage{listings}

\usepackage{color}

\usepackage{textcomp}
\usepackage[table]{xcolor}
\usepackage{multirow}
\usepackage{amsthm}

\usepackage{array}

\usepackage[cmex10]{amsmath}

\theoremstyle{definition}
\newtheorem{definition}{Definition}

\usepackage{amssymb}

\usepackage{mathtools}
\usepackage{mathrsfs}
\DeclareCaptionStyle{ruled}{labelfont=normalfont,labelsep=colon,strut=off} 
\floatstyle{ruled}
\newfloat{listing}{tb}{lst}{}
\floatname{listing}{Listing}
\nocopyright 

\title{
Learning to Solve Optimization Problems with Hard Linear Constraints
}
\author{
    Meiyi Li\textsuperscript{\rm 1}, 
    Soheil Kolouri\textsuperscript{\rm 2},
    Javad Mohammadi\textsuperscript{\rm 1}  
}
\affiliations{
    \textsuperscript{\rm 1}The University of Texas at Austin\\

    \textsuperscript{\rm 2} Vanderbilt University\\
   
}

\usepackage{bibentry}

\begin{document}
\newcommand{\LOOP}{$\mathcal{LOOP}$}
\newcommand{\LOOPLC}{$\mathcal{LOOP-LC}$}
\maketitle

\begin{abstract}
Constrained optimization problems appear in a wide variety of challenging real-world problems, where constraints often capture the physics of the underlying system. Classic methods for solving these problems rely on iterative algorithms that explore the feasible domain in the search for the best solution. These iterative methods are often the computational bottleneck in decision-making and adversely impact time-sensitive applications. Recently, neural approximators have shown promise as a replacement for the iterative solvers that can output the optimal solution in a single feed-forward providing rapid solutions to optimization problems. However, enforcing constraints through neural networks remains an open challenge. This paper develops a neural approximator that maps the inputs to an optimization problem with hard linear constraints to a feasible solution that is nearly optimal. Our proposed approach consists of four main steps: 1) reducing the original problem to optimization on a set of independent variables, 2) finding a gauge function that maps the $\ell_\infty$-norm unit ball to the feasible set of the reduced problem, 3) learning a neural approximator that maps the optimization’s inputs to an optimal point in the  $\ell_\infty$-norm unit ball, and 4) find the values of the dependent variables from the independent variable and recover the solution to the original problem. We can guarantee hard feasibility through this sequence of steps. Unlike the current learning-assisted solutions, our method is free of parameter-tuning and removes iterations altogether. We demonstrate the performance of our proposed method in quadratic programming in the context of the optimal power dispatch (critical to the resiliency of our electric grid) and a constrained non-convex optimization in the context of image registration problems. Our results support our theoretical findings and demonstrate superior performance in terms of computational time, optimality, and the feasibility of the solution compared to existing approaches.

\end{abstract}

\section{Introduction}

\subsection{Motivation}
 
Constrained optimization problems are prevalent in computer sciences and engineering, appearing in various applications. Today's optimization solvers employ iterative solvers that primarily leverage first- and second-order techniques (such as (sub)gradient ascent/descent, conjugate gradients, and simplex basis updating methods) to find the optimal solution. These algorithms often provide theoretical convergence guarantees, which is desirable. However, the iterative nature of these solutions increases calculation time and limits their applicability in time-sensitive applications.
Many practical setups require solving instances of the same problem repeatedly. Another drawback of existing solutions is that their performance does not improve regardless of how often they deal with the same problem. 
Furthermore, the availability of algorithms to handle constrained optimization problems is highly dependent on problem structure, which ranges from problems that can be solved quickly, i.e., linear programming, to problems that have yet to be solved efficiently, e.g., non-convex problems.

The recent interest in using machine learning to improve the efficiency of optimization procedures is fueled by the potential to overcome the discussed shortcomings. Leveraging neural networks can speed up the search process and reduce the number of iterations required to find optimal solutions. The performance of neural approximators can also continually improve as they face more optimization problems.

\subsection{Related works}
\subsubsection{Leveraging deep learning to improve the optimization process of unconstrained problems:}

One of the classical applications of machine learning in optimization has been predicting hyper-parameters, e.g., learning
rate (\citet{yu1995dynamic}), momentum decay  (\citet{smith2018disciplined}), Lagrangian multipliers (\citet{boyd2011distributed}), etc., to enhance the optimization process. 
More recently, Learning to Optimize (L2O) approach aims to replace the engineered traditional optimizers with learnable ones  (\citet{amos2022tutorial,chen2021learning-UT,amos2022meta}). While L2O approaches considerably reduce the total number of iterations required to solve an optimization process, they come short of eliminating iterations altogether.  

Many recent works have focused on replacing the optimization algorithm with a parametric function that directly maps the 
optimization’s input data to the optimal parameters. For instance, the Learning to Optimize the Optimization Process method proposed by \citet{liu2022teaching}
showcases promising results by removing iterations and optimizing the optimization process over time and through different problems. While these iteration-free methods can handle a wide range of unconstrained optimization problems, they often struggle with dealing with constrained problems. Put differently, today's neural approximators (such as the method proposed by \citet{liu2022teaching}) are effective in finding high-quality (near-optimal) solutions but have limited capabilities in finding feasible solutions.


\subsubsection{Using penalty terms to handle constrained optimization problems}

 Incorporating a penalty term to constrain the output of neural approximators is an intuitive strategy. Often the $\ell_2$-norm term enforces equality constraints while penalizing square-of-maximum violation deals with inequality constraints  (\citet{liu2022teaching,pan2020deepopf,liu2022topology}). Alternative penalty terms include (i) the difference between the output and its projection on the constraint set  (\citet{10.1145/3447555.3464874}),  (ii) the discrepancy between the output and its projection on a ball centered 
at the optimal solution  (\citet{detassis2020teaching}), (iii) the status deviation of inequality constraints (whether the optimal solution satisfies the inequality constraints)   (\citet{hasan2021hybrid,zhang2021convex}), and (iv) the violation of KKT conditions  (\citet{nellikkath2022physics, chen2022learning}). 

One of the main drawbacks of enforcing constraints through penalty terms is the need for parameter tuning. The weights of penalty terms are usually determined heuristically, and the performance of these methods is highly sensitive to these parameters. To address this challenge, \citeauthor{pan2020deepopf} presents a design method for proper penalty parameter selection, considering the scalability of the problem.  \citet{Fioretto2020PredictingAO}, and  \citet{tran2020differentially} combine the Lagrangian dual approaches and deep learning to solve the optimal power flow problems with constraints. Compared to methods with user-selected penalty parameters, the Lagrangian dual approach automatically modifies penalty settings during training and produces more reliable results.

\subsubsection{Deep learning for optimization with hard boundaries}

Penalty approaches provide a soft boundary on the output because infeasibility is merely punished rather than eliminated. These methods require a trade-off between optimality and feasibility, with the worst-case scenario being that neither is fulfilled. Approximating solutions to optimization problems with hard restrictions are also explored in several recent works. Some works adopt the  ``projected output" to ensure feasibility.  \citet{detassis2020teaching} have developed an iterative strategy for modifying the objective function to match model predictions more closely. They use an external solver to maintain feasibility. However,  \citet{frerix2020homogeneous} show that projection-based methods might not provide enough information to find the optimal solution since only a limited number of points on the boundary are accessible (due to projection).

Other methods restructure neural networks to ensure feasibility. For example, the double description approach is employed  to cope with linear inequality constraints in \citet{frerix2020homogeneous}. The algorithm constructs a polyhedron-like feasible set iteratively, then optimizes over this constructed feasible set. \citet{hruby2022learning} develop a solver for similar equation-constrained problems based on the homotopy continuation method.  \citet{donti2021dc3} utilize a correction process to solve the feasible problem using gradient descent in each step.  While these methods satisfy hard-inequality constraints, they require an iterative process for training and testing, which goes against the goal of using deep learning models to replace iterative optimization progress.

 \citet{donti2021dc3} and \citet{pan2020deepopf} employ the notion of variable elimination to impose equality constraints when only a subset of the variables is generated in the procedures, and the remaining variables can be inferred using the equality constraints. The variable elimination method is iteration free; hence, it directly produces a feasible solution with respect to equality constraints each time an action is executed.


\subsection{Contributions}
This paper develops a trainable parametric function that directly maps the input to the predicted solution of linearly constrained optimization problems. The proposed method achieves optimality while enforcing hard feasibility. Rather than solving the
original linearly constrained optimization problem directly,
we reformulate and relax it to 
an equivalent optimization problem in the $\ell_\infty$-norm unit ball and train a neural network to find
an optimal solution in the $\ell_\infty$-norm unit ball space. Inspired by \citet{tabas2021computationally}, we construct a one-to-one mapping
to transfer the optimal solution from the $\ell_\infty$-norm unit ball to
the feasible space of the original constraint set. The proposed method is iteration free and enforces equality and inequality
constraints of the original linearly
constrained problem.

\section{Problem Formulation}
In this section, we introduce the notations and problem formulations. Let us consider the following linear-constraint optimization problem:

\begin{align}
    &\min f(\mathbf{u},\mathbf{x})\nonumber\\
\texttt{s.t.}~~~&    \mathbf{A}_{\texttt{eq}}\mathbf{u}+\mathbf{B}_{\texttt{eq}}\mathbf{x}+\mathbf{b}_{\texttt{eq}}=\mathbf{0}\nonumber\\
&    \mathbf{A}_{\texttt{ineq}}\mathbf{u}+\mathbf{B}_{\texttt{ineq}}\mathbf{x}+\mathbf{b}_{\texttt{ineq}}\leq\mathbf{0}
\label{eq:problem1}
\end{align}
Here, $\mathbf{u}\in \mathbb{R}^{N_\texttt{opt}}$ denotes the vector of optimization variables whereas
$\mathbf{x}\in \mathbb{R}^{N_\texttt{inp}}$ represents the input vector. Also, $N_\texttt{opt}$ and $N_\texttt{inp}$ represent dimensions of optimization variables and input vectors.
Moreover, $f(\mathbf{u},\mathbf{x})$ refers to any convex or non-convex objective function. 
We assume that \eqref{eq:problem1} is an under-determined problem, i.e., $\textup{rank}(\mathbf{A}_{\texttt{eq}})=N_{\texttt{eq}}<N_{\texttt{opt}}$. To simplify the notation, we will refer to the optimal solution and constraint set of problem \eqref{eq:problem1} as $\mathbf{u}^*$ and $\mathcal{S}$, respectively. Given that $\mathbf{u}$ is bounded in most practical settings, $\mathcal{S}$ is considered a bounded set. Therefore \eqref{eq:problem1} can be presented by the following abstract form,
\begin{equation}
    \min f(\mathbf{u},\mathbf{x}) \hspace{.3cm} \texttt{s.t.} \hspace{.3cm} \mathbf{u} \in \mathcal{S}(\mathbf{x}) \label{eq:problem1-abst.}
\end{equation}


\begin{figure*}[htbp]
\centering
\setlength{\abovecaptionskip}{0.cm}
\includegraphics[width=1.7\columnwidth]{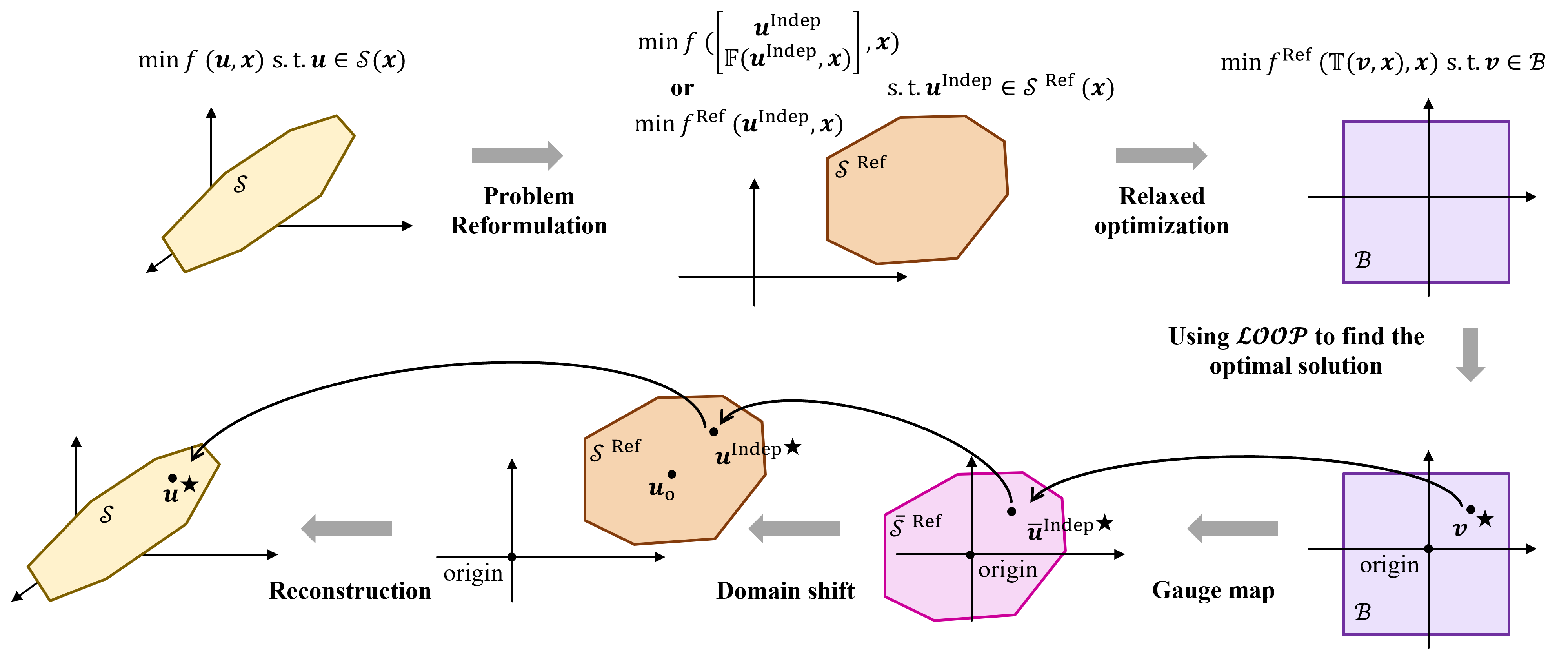}
\caption{structure of the proposed \LOOPLC~model.}
\label{f:flow}
\end{figure*}
Similar to the framework proposed by \citet{liu2022teaching}, we replace the classic iterative solvers with a trainable parametric function $\xi_{\theta}$ that directly maps the input of the optimization problem to the optimal parameters in a single feed-forward. By bypassing the traditional iterative solutions,  the method overcomes one of the significant optimization bottlenecks enabling near real-time optimization in a wide range of critical applications.

Note, problem \eqref{eq:problem1} is a constrained optimization problem. Hence, the feasibility of $\xi_{\theta}$'s output should be ensured. It is straightforward to use activation functions to guarantee feasibility when $\mathcal{S}$ constitutes an $\ell_\infty$-norm ball. 
However, it is challenging to utilize neural approximators to solve \eqref{eq:problem1-abst.} where $\mathcal{S}$ includes coupled constraints and variables (see  
appendix for more details).
This paper extends the prior works (e.g., \citet{liu2022teaching,amos2022tutorial}) for solving unconstrained problems to solve linearly constrained optimization problems (referred to as \LOOPLC, Learning to Optimize the Optimization Process with Linear Constraints). In what follows, we will first introduce the basics of \LOOPLC~method and provide theoretical guarantees to justify the feasibility of the resulting solution. Later, We will showcase the performance of \LOOPLC~in quadratic programming in the context of the optimal power dispatch (critical to the resiliency of our electric grid) and a constrained non-convex optimization in the context of image registration problems.
 

\section{Proposed Architecture}
\label{Proposed Architecture}
The rest of this section is dedicated to presenting an abstract overview of our proposed framework, \LOOPLC.

\noindent (i) \textit{Optimization reformulation}: We will adopt a variable elimination technique \shortcite{donti2021dc3} to reformulate
\eqref{eq:problem1} as a reduced-dimension optimization problem with only inequalities. This approach first decomposes the set of optimization variables $\mathbf{u}$ to independent ($\mathbf{u}^\texttt{Indep}$) and dependent ($\mathbf{u}^\texttt{Dep}$) parts. Then uses equality constraints to find relationships between $\mathbf{u}^\texttt{Indep}$ and  $\mathbf{u}^\texttt{Dep}$, i.e., $ \mathbf{u}^{\texttt{Dep}}=\mathbb{F}(\mathbf{u}^{\texttt{Indep}},\mathbf{x})$.
The reformulated problem will be referred to as, 
\begin{equation}
    \min f^{\texttt{Ref}}(\mathbf{u^\texttt{Indep}},\mathbf{x}) \hspace{.3cm} \texttt{s.t.} \hspace{.3cm} \mathbf{u}^\texttt{Indep} \in \mathcal{S}^{\texttt{Ref}}(\mathbf{x}) \label{eq:problem1-abst.reform}
\end{equation}
where $f^{\texttt{Ref}}$ is a reformulation of $f$ that only depends on $\mathbf{u}^\texttt{Indep}$.


\noindent (ii) \textit{Utilizing neural network to optimize \eqref{eq:problem1-abst.reform}}. 
Instead of solving the original problem directly, we train a neural network to find the optimal solution in the $\ell_\infty$-norm unit ball, $\mathcal{B}$. 


\begin{equation}
    \min f^{\texttt{Ref}}(\mathbb{T}(\mathbf{v},\mathbf{x}),\mathbf{x}) \hspace{.3cm} \texttt{s.t.} \hspace{.3cm} \mathbf{v} \in \mathcal{B} \label{eq:problem1-abst.unitball}
\end{equation}

Layers of the neural network will ensure that the resulting $\mathbf{v}^{\star}$ will stay within $\mathcal{B}$. Later we will provide a one-to-one mapping to transfer the resulting solutions from $\ell_\infty$-norm unit ball to the feasible space of constraint set $\mathcal{S}^{\texttt{Ref}}$. This mapping will be denoted as $\mathbf{u}^\texttt{Indep}=\mathbb{T}(\mathbf{v},\mathbf{x})$.

\noindent (iii) \textit{Ensuring feasibility by finding $\mathbb{T}$}:
We use the gauge map \cite{tabas2021computationally} to build a one-to-one mapping between $\mathcal{B}$ (i.e., $\ell_\infty$-norm unit ball) and $\mathcal{S}^{\texttt{Ref}}$ (i.e., feasible domain of the reformulated problem) spaces. The gauge map requires the destination space to encompass the origin as an interior point \cite{blanchini2008set}.
Thus, rather than directly mapping the $\ell_\infty$-norm unit ball into the desired feasible domain $\mathcal{S}^{\texttt{Ref}}$, we first shift the desired domain by one of its interior points $\mathbf{u}_{\texttt{o}}$ to construct an ``intermediate domain" $\mathcal{\bar{S}}^{\texttt{Ref}}$. The intermediate domain shares the same geometric properties with $\mathcal{S}^{\texttt{Ref}}$ but contains the origin as an interior point. In a nutshell, this process enables leveraging the gauge function to map the $\ell_\infty$-norm unit ball into an ``intermediate domain" and then shift the intermediate domain to the desired domain, $\mathcal{S}^{\texttt{Ref}}$.  

\noindent (iv) \noindent\textit{Equality completion}: 
The previous step tackles the reduced-dimension optimization problem (i.e., formulation \eqref{eq:problem1-abst.reform}), hence, it finds optimal values for a subset of optimization variables (i.e., ${\mathbf{u}^\texttt{Indep}}^{\star}$).
The remaining variables  ${\mathbf{u}^\texttt{Dep}}^{\star}$, will be determined by using the algebraic relationship $\mathbb{F}$ between optimization variables that are given by equality constraints of \eqref{eq:problem1-abst.}.

\begin{figure}[htbp]
\centering
\setlength{\abovecaptionskip}{0.cm}
\includegraphics[width=1\columnwidth]{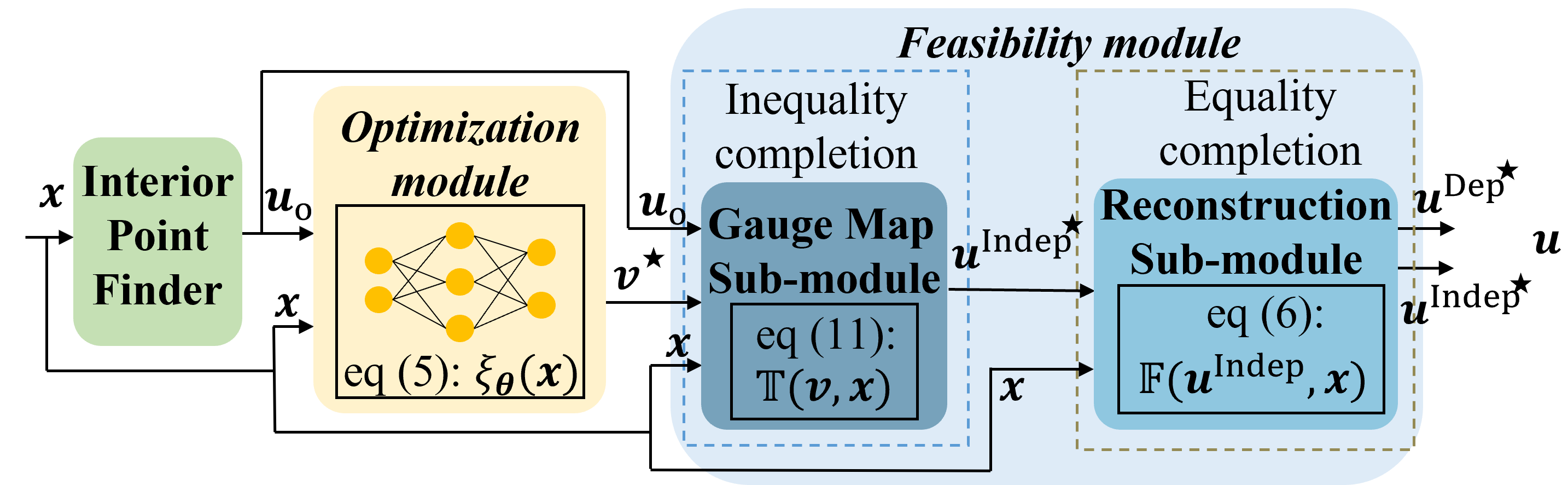}
\caption{The \LOOPLC~ framework is composed of interior point finder, optimization, and feasibility modules.}\centering
\label{f:basic}
\end{figure}

The forthcoming sections present the details of the steps presented in this section. The modular architecture of \LOOPLC, depicted in Figure \ref{f:basic}, ensures achieving a high-quality (near-optimal) feasible solution.


\section{Optimization Module}
\label{Optimization module}
As stated in section \ref{Proposed Architecture}, we incorporate a neural network
to learn a high-quality solution to problem \eqref{eq:problem1-abst.unitball}. Note that the gauge map sub-module (that will be introduced later) 
takes in the interior point $\mathbf{u}_\texttt{o}(\in \mathcal{S}^{\texttt{Ref}})$, hence the \textbf{input}s of the optimality module include both $\mathbf{u}_\texttt{o}$ and $\mathbf{x}$. Let $\boldsymbol{\theta}$ denote the weights of the neural network. 
The \textbf{output} of the optimality module is prediction $\mathbf{v}$ that lies in the $\ell_\infty$-norm unit ball (i.e., $\mathbf{v}^\star$):

\begin{align}
    \mathbf{v}^\star={\xi}_{\boldsymbol{\theta}}(\mathbf{x},\mathbf{u}_\texttt{o})
\end{align}

Choosing the proper activation functions, e.g., the hyperbolic tangent function, ensures that the resulting $\mathbf{v}$ stays in the feasible range of $\ell_\infty$-norm unit ball, i.e., $[-1,1]$.
Later, $\mathbf{v}^\star$ will pass through the feasibility module to generate $\mathbf{u}^\star$.

\begin{figure}[htbp]
\centering
\setlength{\abovecaptionskip}{0.cm}
\includegraphics[width=0.8\columnwidth]{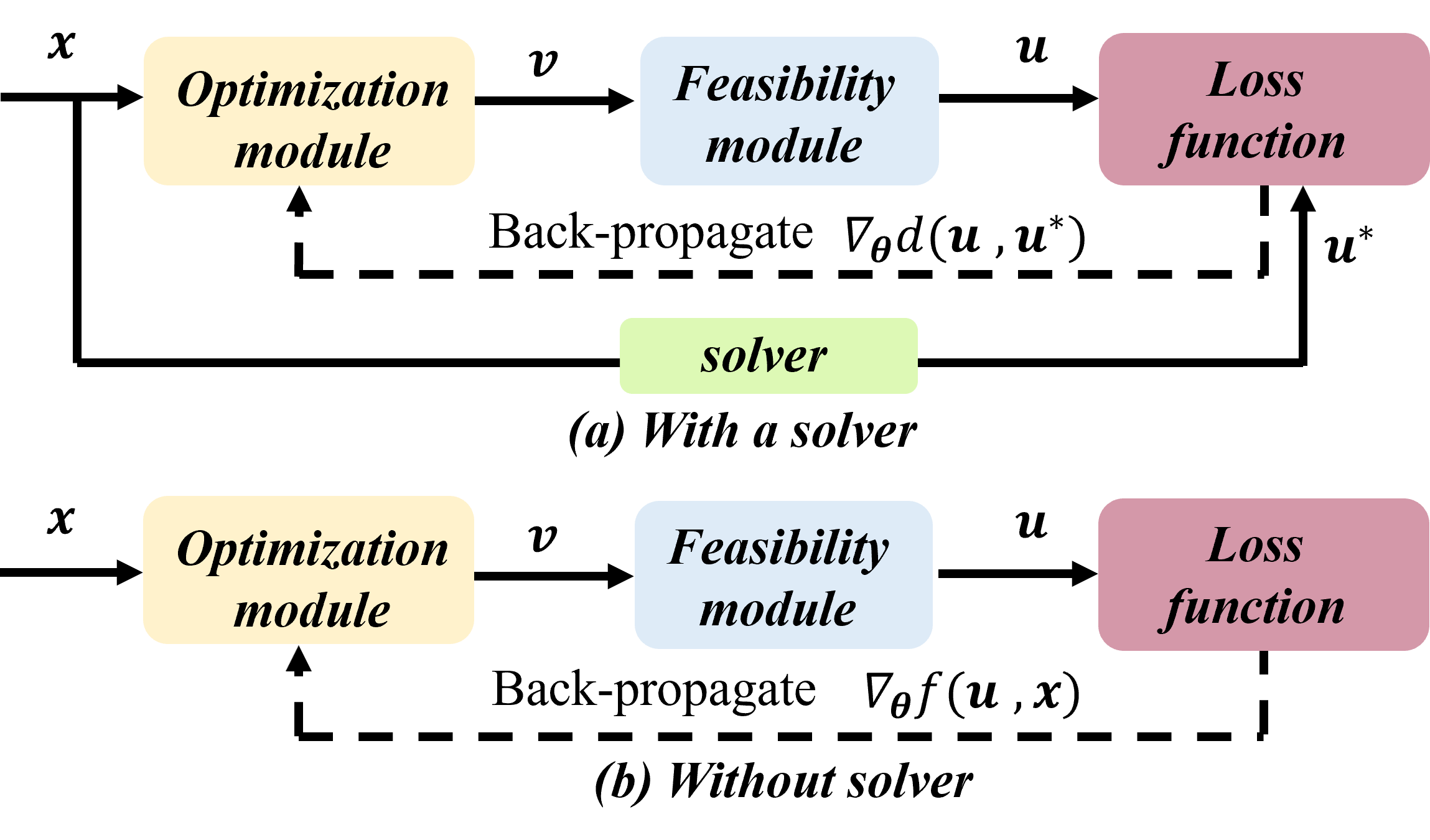}
\caption{Two training approaches of \LOOPLC~model}
\label{f:train}
\end{figure}

The optimality module uses two training approaches, as illustrated in Figure \ref{f:train}:
1) with a solver in the loop, and 2) without a solver in the loop, i.e., directly minimizing the objective function. Assume there are $N$ input data points indexed as $\mathbf{x}^{(i)}$, the respective output denoted as $\mathbf{u}^{(i)}$.
The loss function with a solver in the loop is a discrepancy/distance function $d:\mathbb{R}^{N_\texttt{opt}}\times \mathbb{R}^{N_\texttt{opt}}\rightarrow \mathbb{R}_{+}$ defined in $\mathbb{R}^{N_\texttt{opt}}$ and
compares the difference between the output of the \LOOPLC~ model and the optimal solution calculated using commercial solvers, i.e. $L=\sum_{i=1}^{N}d( \mathbf{u}^{(i)},\mathbf{u}^{(i)*})$. Without a solver, the loss function is just  the expected value of the objective function, i.e., $L=\sum_{i=1}^{N}f(\mathbf{u}^{(i)},\mathbf{x}^{(i)})$.


\section{Feasibility Module}
\label{Feasibility module}
While the optimality module section provides a high-quality solution, it does not necessarily produce a feasible solution.
To this end, the feasibility module will first map $\mathbf{v}^\star$ onto the desired feasible domain and then compute a full-dimensional output $\mathbf{u}^\star$. In this section, we will present how this module enforces feasibility through equality completion and inequality satisfaction.

\subsection{Equality completion}

This sub-module reconstruct the equality equations of \eqref{eq:problem1}. 
We first divide the elements in $\mathbf{u}$ into two groups: $(N_{\texttt{opt}}\!-\!N_{\texttt{eq}})$ independent parameters and $N_{\texttt{eq}}$ dependent parameters. Dependent parameters $\mathbf{u}^{\texttt{Dep}} \in \mathbb{R}^{N_{\texttt{eq}}}$ are defined by all the equality constraints in problem \eqref{eq:problem1} whereas independent parameters $\mathbf{u}^{\texttt{Indep}} \in \mathbb{R}^{(N_{\texttt{opt}}-N_{\texttt{eq}})}$.

\textbf{Claim:} Let us define function $\mathbb{F}$ as $\mathbb{F}:\mathbb{R}^{(N_{\texttt{opt}}-N_{\texttt{eq}})}\rightarrow \mathbb{R}^{N_{\texttt{eq}}}$ s.t. $\mathbf{u}^{\texttt{Dep}}=\mathbb{F}(\mathbf{u}^{\texttt{Indep}},\mathbf{x})$ where \small$\mathbf{A}_{\texttt{eq}}\begin{bmatrix}
\mathbf{u}^{\texttt{Indep}}\\ 
\mathbf{u}^{\texttt{Dep}}
\end{bmatrix}=-\mathbf{B}_{\texttt{eq}}\mathbf{x}-\mathbf{b}_{\texttt{eq}}$\normalsize.
Then, there exists such an $\mathbb{F}$ for a linear-constraint set.

\textbf{Justification:} Let us select $N_{\texttt{eq}}$ linearly independent columns in $\mathbf{A}_{\texttt{eq}}$ and group them into $\mathbf{A}_{\texttt{eq}}^{\texttt{Dep}}$. The other columns form $\mathbf{A}_{\texttt{eq}}^{\texttt{Indep}}$. Then we have 
\small$\mathbf{A}_{\texttt{eq}}^{\texttt{Indep}}\mathbf{u}^{\texttt{Indep}}+\mathbf{A}_{\texttt{eq}}^{\texttt{Dep}}\mathbf{u}^{\texttt{Dep}}+\mathbf{B}_{\texttt{eq}}\mathbf{x}+\mathbf{b}_{\texttt{eq}}=\mathbf{0}$\normalsize. That is:
\begin{align}
    &\mathbf{u}^{\texttt{Dep}}=\mathbb{F}(\mathbf{u}^{\texttt{Indep}},\mathbf{x})\nonumber\\
    &=-\mathbf{A}_{\texttt{eq}}^{\texttt{Dep}^{-1}}\mathbf{A}_{\texttt{eq}}^{\texttt{Indep}}\mathbf{u}^{\texttt{Indep}}-\mathbf{A}_{\texttt{eq}}^{\texttt{Dep}^{-1}}(\mathbf{B}_{\texttt{eq}}\mathbf{x}+\mathbf{b}_{\texttt{eq}})\label{eq:dep_ind}
\end{align}

By incorporating reconstruction function $\mathbb{F}$, problem \eqref{eq:problem1} changes to minimizing \small$f(\begin{bmatrix}
\mathbf{u}^{\texttt{Indep}}\\ 
\mathbb{F}(\mathbf{u}^{\texttt{Indep}},\mathbf{x})
\end{bmatrix},\mathbf{x})$\normalsize with merely inequality constraints \small$\mathbf{A}_{\texttt{ineq}}\begin{bmatrix}
\mathbf{u}^{\texttt{Indep}}\\ 
\mathbb{F}(\mathbf{u}^{\texttt{Indep}},\mathbf{x})
\end{bmatrix}+\mathbf{B}_{\texttt{ineq}}\mathbf{x}+\mathbf{b}_{\texttt{ineq}}\leq\mathbf{0}$\normalsize. We rewrite  $\mathbf{A}_{\texttt{ineq}}$ as $\begin{bmatrix}
\mathbf{A}_{\texttt{ineq}}^{\texttt{Indep}} & \mathbf{A}_{\texttt{ineq}}^{\texttt{Dep}}
\end{bmatrix}$ in accordance with $\mathbf{u}^{\texttt{Indep}}$ and $\mathbf{u}^{\texttt{Dep}}$. Then, according to \eqref{eq:dep_ind} and \eqref{eq:problem1}, the inequality constraints can be written as: \small$ ( \mathbf{A}_{\texttt{ineq}}^{\texttt{Indep}}-\mathbf{A}_{\texttt{ineq}}^{\texttt{Dep}} \mathbf{A}_{\texttt{eq}}^{\texttt{Dep}^{-1}}\mathbf{A}_{\texttt{eq}}^{\texttt{Indep}})\mathbf{u}^{\texttt{Indep}}-\mathbf{A}_{\texttt{ineq}}^{\texttt{Dep}}\mathbf{A}_{\texttt{eq}}^{\texttt{Dep}^{-1}}(\mathbf{B}_{\texttt{eq}}\mathbf{x}+\mathbf{b}_{\texttt{eq}})+\mathbf{B}_{\texttt{ineq}}\mathbf{x}+\mathbf{b}_{\texttt{ineq}}\leq\mathbf{0}$\normalsize. As the next step, let us rewrite problem \eqref{eq:problem1} as a reduced-dimension optimization problem with only inequalities, as \eqref{eq:problem1-abst.reform}. This means that \small$\mathbf{A}= \mathbf{A}_{\texttt{ineq}}^{\texttt{Indep}}-\mathbf{A}_{\texttt{ineq}}^{\texttt{Dep}} \mathbf{A}_{\texttt{eq}}^{\texttt{Dep}^{-1}}\mathbf{A}_{\texttt{eq}}^{\texttt{Indep}}$, $\mathbf{B}=\mathbf{B}_{\texttt{ineq}}-$ $\mathbf{A}_{\texttt{ineq}}^{\texttt{Dep}}\mathbf{A}_{\texttt{eq}}^{\texttt{Dep}^{-1}}\mathbf{B}_{\texttt{eq}}$, $\mathbf{b}=\mathbf{b}_{\texttt{ineq}}-\mathbf{A}_{\texttt{ineq}}^{\texttt{Dep}}\mathbf{A}_{\texttt{eq}}^{\texttt{Dep}^{-1}}\mathbf{b}_{\texttt{eq}}$\normalsize. That is:
 
 \begin{align}
   \mathcal{S}^{\texttt{Ref}}=\left \{ \mathbf{u}^{\texttt{Indep}}|\mathbf{A}\mathbf{u}^{\texttt{Indep}}+\mathbf{B}\mathbf{x}+\mathbf{b} \leq\mathbf{0}\right \} \\
   f^{\texttt{Ref}}(\mathbf{u}^{\texttt{Indep}},\mathbf{x})=f(\begin{bmatrix}
\mathbf{u}^{\texttt{Indep}}\\ 
\mathbb{F}(\mathbf{u}^{\texttt{Indep}},\mathbf{x})
\end{bmatrix},\mathbf{x})
 \end{align}

Thus, we can solve \eqref{eq:problem1-abst.reform}
and later reconstruct the remaining parameters according to \eqref{eq:dep_ind}. Therefore, the reconstruction sub-module takes in independent parameters $\mathbf{u}^{\texttt{Indep}}$ and $\mathbf{x}$ and outputs $\mathbf{u}^{\texttt{Dep}}$. The existence of $\mathbb{F}$ guarantees that the resulting optimization solution $\mathbf{u}^\star=\begin{bmatrix}
{\mathbf{u}^{\texttt{Indep}}}^\star\\ 
{\mathbf{u}^{\texttt{Dep}}}^\star
\end{bmatrix}$ satisfy the equality constraints of \eqref{eq:problem1}

\subsection{Inequality completion}

In order to enforce inequality constraints in \eqref{eq:problem1-abst.reform}, we incorporate the gauge map sub-module, which is based on the Minkowski function defined below. 


\begin{definition}[Minkowski function]
Given a convex and compact set $\mathcal{C}\subset\mathbb{R}^{n}$, assume the origin belongs to the algebraic interior of $\mathcal{C}$ and $\mathbf{c}\in \mathcal{C}$.
The Minkowski function associated with $\mathcal{C}$ is defined by $\varphi_{\mathcal{C}}(\mathbf{c})=\inf\left \{ r>0 :\mathbf{c}\in r\mathcal{C}\right \}.$

\end{definition}

The space $\mathcal{C}$ is a polytope (encompassing the origin) and is defined as $\mathcal{C}=\left \{ \mathbf{c}\in \mathbb{R}^{n} |\mathbf{F}^j\mathbf{c}\leq g^j, j=1...m\right \}$.
Here, $\mathbf{F}^j$ is a $1\times n$ row vector. The Minkowski function associated with $\mathcal{C}$ is defined as:
\begin{align}
  \varphi_{\mathcal{C}}(\mathbf{c})=\underset{j}{\max}\{\frac{\mathbf{F}^j\mathbf{c}}{g^j}\}  
\end{align}

The Minkowski function allows ``translating" specific geometric properties of a subset to a (particular) algebraic property of another subset. The ``translation" is enabled by the gauge map.

\begin{figure}[htbp]
\centering
\setlength{\abovecaptionskip}{0.2cm}
\includegraphics[width=0.5\columnwidth]{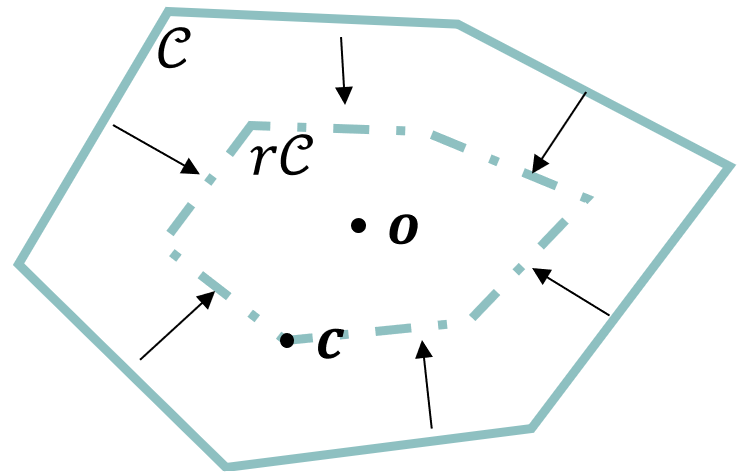}
\caption{The sub-level sets of a Minkowski function are achieved by linearly scaling the set $\mathcal{C}$. Specifically, any point $\mathbf{c}$ could be referred to by the distance to the origin $\varphi_{\mathcal{C}}(\mathbf{c})$ and the direction $\mathbf{c}/\varphi_{\mathcal{C}}(\mathbf{c})$.}
\label{f:scale}
\end{figure}

\begin{definition}[Gauge mapping function]
Let us consider two convex and compact sets $\mathcal{C}\subset\mathbb{R}^{n}$ and $\mathcal{\bar{C}}\subset\mathbb{R}^{n}$. Let us assume that the origin belongs to the algebraic interior of both $\mathcal{C}$ and $\mathcal{\bar{C}}$. The gauge map $G: \mathcal{C}\rightarrow \mathcal{\bar{C}}$ is a bijection function defined as $\mathbf{\bar{c}}=G(\mathbf{c},\mathcal{C},\mathcal{\bar{C}})=\frac{\varphi_{\mathcal{C}}(\mathbf{c})}{\varphi_{\mathcal{\bar{C}}}(\mathbf{c})}\mathbf{c}$. Here, $\mathbf{c}\in \mathcal{C}$ and $\mathbf{\bar{c}}\in \mathcal{\bar{C}}$.

\end{definition}
 
This property means a feasible range with a simple geometric shape (such as $\ell_\infty$-norm unit ball $\mathcal{B}$) can be translated to a complex feasible range (such as  $\mathcal{\bar{S}}^{\texttt{Ref}}$). Since the gauge map function provides a one-to-one mapping, choosing a point in $\mathcal{B}$  is equivalent to choosing a point in  $\mathcal{\bar{S}}^{\texttt{Ref}}$.  

As shown in Figure \ref{f:scale}, the gauge map function is based on the concept of an absorbing set that can be deflated in accordance with the origin. To apply the gauge map function, however, we must
temporarily “shift” the desired feasible domain $\mathcal{S}^{\texttt{Ref}}$ by one of its interior points $\mathbf{u}_{\texttt{o}}$ to make it 
a set $\mathcal{\bar{S}}^{\texttt{Ref}}$
that contains the origin as an interior point. Thus,
\begin{align}
    \mathcal{\bar{S}}^{\texttt{Ref}}=\{\mathbf{\bar{u}}^{\texttt{Indep}}|(\mathbf{\bar{u}}^{\texttt{Indep}}+\mathbf{u}_{\texttt{o}})\in \mathcal{S}^{\texttt{Ref}}\}
\end{align}

The set $\mathcal{\bar{S}}^{\texttt{Ref}}$ serves as a bridge connecting the $\ell_\infty$-norm unit ball $\mathcal{B}$ and the desired feasible domain $\mathcal{S}^{\texttt{Ref}}$. We use the gauge map to translate $\mathbf{v}^{\star}$ into a $\mathbf{\bar{u}}^{{\texttt{Indep}}^{\star}}$ in $\mathcal{\bar{S}}^{\texttt{Ref}}$ and then shift it to $\mathbf{u}^{{\texttt{Indep}}^{\star}}$ in $\mathcal{S}^{\texttt{Ref}}$. Put differently, 
\begin{align}
    \mathbf{u}^{{\texttt{Indep}}^{\star}}\!\!=\!\!\mathbb{T}({\mathbf{v}}^{\star},\mathbf{x})
    \!\!=\!\!\mathbf{\bar{u}}^{{\texttt{Indep}}^{\star}}\!\!+\!\!\mathbf{u}_{\texttt{o}}\!\!=\!\!
    \frac{\varphi_{\mathcal{B}}(\mathbf{v}^{\star})}{\varphi_{\mathcal{\bar{S}}^{\texttt{Ref}}}(\mathbf{v}^{\star})}\mathbf{v}^{\star}\!\!+\!\!\mathbf{u}_{\texttt{o}}\label{eq:h}
\end{align}

Therefore, the \textbf{inputs} of the gauge map sub-module are $\mathbf{x}$, $\mathbf{v}^\star$ and an interior point $\mathbf{u}_{\texttt{o}}$. The  \textbf{output} is independent parameters ${\mathbf{u}^{\texttt{Indep}}}^\star$.

All in all, given any $\mathbf{v}^\star$ in the $\ell_\infty$-norm unit ball, the feasibility module first produces a reduced-size solution ${\mathbf{u}^{\texttt{Indep}}}^\star$ and then expands it to a full-dimension solution $\mathbf{u}^\star$. The gauge map and reconstruction functions will enforce 
both the equality constraints and inequality constraints. 

\section{Interior Point Finder}\label{Interior}
As discussed in the optimization and feasibility modules, we need to use an interior point in $\mathcal{S}^{\texttt{Ref}}$ to 
construct the intermediate domain $\mathcal{\bar{S}}^{\texttt{Ref}}$. The difficulty of finding interior points stems from the fact that $\mathcal{S}^{\texttt{Ref}}$ varies as $\mathbf{x}$ changes. Therefore, an interior point of $\mathcal{S}^{\texttt{Ref}}(\mathbf{x}^{(i)})$ may not be an interior point of $\mathcal{S}^{\texttt{Ref}}(\mathbf{x}^{(j)}), j\neq i$. We start by making an assumption that given any 
$\mathbf{x}^{(i)}$, $i=1...m$, $\exists  \mathcal{S}^{\texttt{int}}\subset \mathcal{S}^{\texttt{Ref}}(\mathbf{x}^{(i)})  $ (\citet{tabas2021computationally}). Then any point in 
$\mathcal{S}^{\texttt{int}}$ is an interior point. This assumption holds when the input $\mathbf{x}$ (or  $\mathcal{S}^{\texttt{Ref}}(\mathbf{x})$) is under small disturbances. In this section, we present an initial artificial problem method to find out an interior point for more general cases. 
The appendix outlines two alternative interior point finders.

Inspired by the implementation of the interior-point method \cite{adler1989implementation}, we first define the following problem using the pseudo-variable $u_a\in \mathbb{R}$,


\begin{align}
    &\min Mu_a\nonumber\\
\texttt{s.t.}~~~&    
    \mathbf{A}\mathbf{u}^{\texttt{Indep}}+\mathbf{B}\mathbf{x}+\mathbf{b} -\mathbf{1}u_a\leq \mathbf{0}
\label{eq:problemif}
\end{align}

Here, $M$ is a large coefficient. $\mathbf{1}$ is an all-one column vector. The solution of \eqref{eq:problemif} is an interior point to $\mathcal{S}^{\texttt{Ref}}$.


Let us note $\begin{bmatrix}
\mathbf{u}^{\texttt{Indep}\blacklozenge}, 
u_a^{\blacklozenge}
\end{bmatrix}^T$ as the solution of \eqref{eq:problemif}. The interior point of $\mathcal{S}^{\texttt{Ref}}$ exists if and only if  $u_a^{\blacklozenge}<0$. Thus, solving problem \eqref{eq:problemif} yields  
an interior point ${\mathbf{u}^{\texttt{Indep}}}^{\blacklozenge} \in \mathcal{S}^{\texttt{Ref}}$ if $u_a^{\blacklozenge}<0$. 

Note, problem \eqref{eq:problemif} is a simple linear programming problem, and it can be solved efficiently using classic solvers. The discussed interior finder method works well when there are not many constraints. However, for large-scale constrained problems, the iterative nature of classical solvers will still be limited, and we postpone it to future work.

\section{Results}

\subsection{Convex problem: DC optimal power flow}

The DC optimal power flow (DCOPF) problem minimizes the cost of procuring electricity in a power grid while respecting the system's limitations. The formulation of the DCOPF problem is given below.


\begin{align}
    \min f(\mathbf{P}_{\texttt{G}})=\sum_{g=1}^{N_{\texttt{G}}} f^{g}(P_{\texttt{G}}^g)\nonumber\\
s.t.\sum_{g=1}^{N_{\texttt{G}}}P_{\texttt{G}}^g=\sum_{i=1}^{N_{\texttt{D}}}P_{\texttt{D}}^i\textup{ , }
\mathbf{\underline{P}_{\texttt{G}}}\leq \mathbf{P_{\texttt{G}}}\leq \mathbf{\overline{P}_{\texttt{G}}}\nonumber\\
\mathbf{{D}_{\texttt{G}}}\mathbf{P_{\texttt{G}}}-\mathbf{{D}_{\texttt{D}}}\mathbf{P_{\texttt{D}}}\leq \mathbf{P_{\texttt{line}}}
\label{eq:dcopf}
\end{align}

Where $\mathbf{P_{\texttt{G}}}= [ P_{\texttt{G}}^1...P_{\texttt{G}}^g...P_{\texttt{G}}^{N_{\texttt{G}}}  ]^T$ refers to the vector of electric power generation,
$\mathbf{P_{\texttt{D}}}= [ P_{\texttt{D}}^1...P_{\texttt{D}}^i...P_{\texttt{D}}^{N_{\texttt{D}}} ]^T$ denote the vector of electric demands. The equality constraint enforces the balance of supply and demand, while the inequality constraint respects the physics of the electric system. 
Here, $\mathbf{{D}_{\texttt{g}}}$ and $\mathbf{{D}_{\texttt{D}}}$ (the so called power transfer distribution factor matrix \cite{ronellenfitsch2016dual}) capture the physics of the electric network.


\textbf{Dataset}: We use the publicly available IEEE 200-bus system data set, available via the MATPOWER (\citet{zimmerman2010matpower}), as the seed information to
generate 200 data points (with a train/test ratio of 1:1). The IEEE 200-bus system is a 200 nodes graph representing a realistic electric grid. This system consists of 200 load nodes and 49 generation nodes. We consider a 10-percentage fluctuation of each load node. 

\begin{figure}[htbp]
\centering
\setlength{\abovecaptionskip}{0.cm}
\includegraphics[width=1\columnwidth]{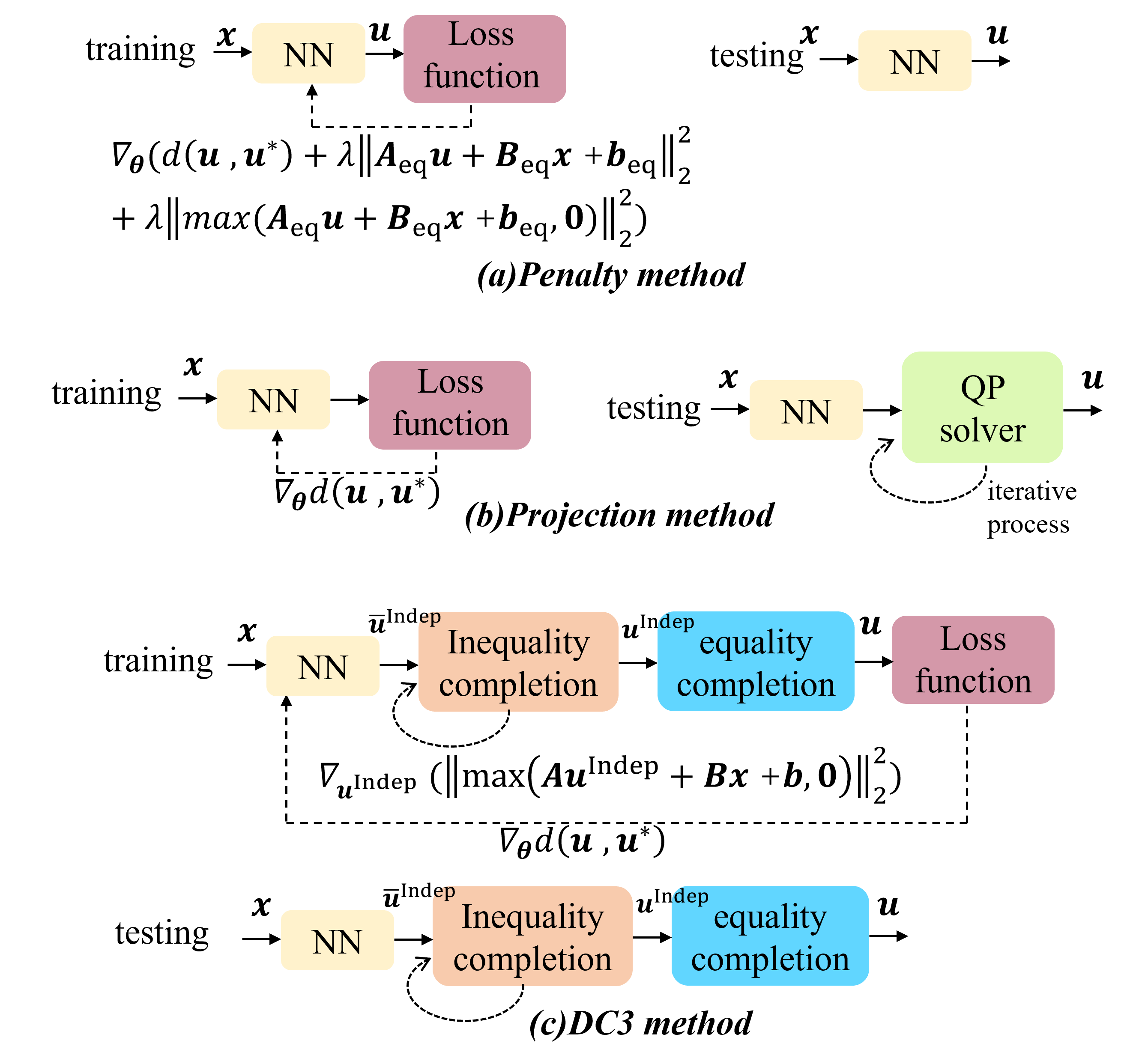}
\caption{We present an illustrative comparison between functionalities of projection, penalty, and DC3
method for solving \eqref{eq:problem1}}
\label{f:other}
\end{figure}

\textbf{Comparison}: We compare \LOOPLC~against the three recent learning-based optimization methods; (i) projection  \cite{10.1145/3447555.3464874}, (ii) penalty  \cite{liu2022teaching}, and (iii) DC3  \cite{donti2021dc3} methods, as well as two well-known commercial solvers (i.e., matpower 7.1 \cite{zimmerman2010matpower} and CVXOPT \cite{vandenberghe2010cvxopt}). The projection method projects the output of the neural network onto the feasible range, while the penalty method adds a $\ell_2$-norm term to the loss function. Moreover, the DC3 method utilizes the $\ell_2$-norm penalty term in the objective function to iteratively enforce the output of the neural networks to satisfy optimization constraints.
These methods are illustratively compared in Figure \ref{f:other}.

\textbf{Parameters}: We use a fixed neural network architecture for projection, penalty, and DC3 methods: fully connected with one
hidden layer of size 16, including ReLU activation. An extra Tanh activation is added to the output layer for \LOOPLC~model.  Different hyperparameters were tuned to maximize performance for each method individually (see Table \ref{t:Parameter}). Since the DC3 method is based on an inner iteration, the performance improves as iteration goes. We let it run till a similar time limit as \LOOPLC~model to facilitate comparison. 

\textbf{Interior point finder}:  $\mathcal{S}^{\texttt{Ref}}$ varies as the electric demand changes; therefore, we use the initial artificial problem method (discussed in Section \ref{Interior}) to find
interior points for the \LOOPLC~method. 


\begin{table}[]
\centering
\caption{Hyperparameters tested for different methods. The final parameter values are identified in bold.}
\label{t:Parameter}
\begin{tabular}{p{0.9cm}p{2.3cm}p{3.8cm}}
\hline
\textbf{Penalty}              & penalty      coefficient               & $10^2$,$10^3$,$\mathbf{10^4}$,$10^5$,$10^6$                \\ \hline
\multirow{3}{*}{\textbf{DC3}} & step size                         & $10^{-2}$,$10^{-3}$,$\mathbf{10^{-4}}$,$10^{-5}$,$10^{-6}$ \\ \cline{2-3} 
                     & inner iteration times for testing & 1,\textbf{3},5,10                                              \\ \cline{2-3} 
                     & inner iteration time for training & 1,\textbf{3},5,10                                                    \\ \hline
\end{tabular}
\end{table}

\begin{table*}[]
\centering
\caption{This table presents the results of using different methods to solve the DCOPF problem. The time is reported as the average per instance in milliseconds. The Optimality gap is measured as $\frac{1}{N}
    \sum_{i=1}^{N}
    \frac{\left \|  \mathbf{u}^{(i)}-\mathbf{u}^{(i)*} \right \|_1}{\left \|  \mathbf{u}^{(i)*} \right \|_1}$. The Feasibility gap is calculated using $\frac{1}{N}$ $\left ( \left \| \max (\mathbf{A}_{\texttt{ineq}}\mathbf{u}^{(i)}+\mathbf{B}_{\texttt{ineq}}\mathbf{x}^{(i)}+\mathbf{b}_{\texttt{ineq}},\mathbf{0})\right \|_1 +\right.$ $ \left. \left \| \mathbf{A}_{\texttt{eq}}\mathbf{u}^{(i)}+\mathbf{B}_{\texttt{eq}}\mathbf{x}^{(i)}+\mathbf{b}_{\texttt{eq}} \right \|_1\right )$}
\label{t:dcopf}
\begin{tabular}{|c|c|c|c|c|}
\hline
                                                           & \textbf{Optimizer} & \textbf{Optimality Gap} & \textbf{Feasibility Gap} & \textbf{Time on CPU(msec)} \\ \hline
\rowcolor[HTML]{FFDBD9} 
\cellcolor[HTML]{EFEFEF}                                   & \multicolumn{1}{p{5.5cm}|}{Matpower \cite{zimmerman2010matpower}}           & 0                       & 0                        & 290.00                     \\ \cline{2-5} 
\rowcolor[HTML]{FFDBD9} 
\multirow{-2}{*}{\cellcolor[HTML]{EFEFEF}Standard Solvers} & \multicolumn{1}{p{5.5cm}|}{CVXOPT \cite{vandenberghe2010cvxopt}}            & 0                       & 0                        & 139.40                     \\ \hline
\rowcolor[HTML]{FFFDC4} 
\cellcolor[HTML]{EFEFEF}                                   & \multicolumn{1}{p{5.5cm}|}{Projection-Based \cite{10.1145/3447555.3464874}}  & 0.00194                 & 0                        & 221.24                     \\ \cline{2-5} 
\rowcolor[HTML]{FFFDC4} 
\cellcolor[HTML]{EFEFEF}                                   & \multicolumn{1}{p{5.5cm}|}{Penalty-Based \cite{liu2022teaching} }    & 0.01524                 & 0.04819                  & 0.16                       \\ \cline{2-5} 
\rowcolor[HTML]{FFFDC4} 
\cellcolor[HTML]{EFEFEF}                                   & \multicolumn{1}{p{5.5cm}|}{DC3 \cite{donti2021dc3}}               & 0.01032                 & 0.04579                  & 0.80                       \\ \cline{2-5} 
\rowcolor[HTML]{B5E8B4} 
\multirow{-4}{*}{\cellcolor[HTML]{EFEFEF}Neural Solvers}   & Our                & 0.00203                 & 0                        & 0.76                       \\ \hline
\end{tabular}\label{DCOPF-results}
\end{table*}
   


\textbf{Results}: Based on Table \ref{DCOPF-results}, the penalty method achieves the best execution time. However, at its core, this method introduces a trade-off between optimality and feasibility. The convergence speed up comes at the cost of an increased feasibility gap.
The same trade-off manifests itself in the DC3 method, meaning that the feasibility gap decreases with more inner iterations, which may adversely impact optimality and solution time. The performance of both the penalty approach and the DC3 method is sensitive to hyperparameters. Poor choices of hyperparameters may lead to divergence of training or testing (for example, the step size of 0.001 for DC3 results in divergence).

Our results show that only \LOOPLC~and projection method satisfy hard feasibility constraints, while \LOOPLC's solution time performance surpasses the projection method by a large margin. 
Specifically, given the interior point, the \LOOPLC~method will be executed three orders of magnitude faster than the projection method.

\subsection{Nonconvex problem: Image registration}
Image registration is a fundamental image analysis problem that aims to optimize the transform function that moves the coordinate system of one image to another (\citet{murez2018image}). Let $\mathcal{X}=[-1,1]^2$ denote the domain of an image, and let us denote $I:\mathcal{X}\rightarrow [0,1]$ as an image defined in this domain. Also, we will refer to the
source image as $I_{\texttt{s}}$ and target image $I_{\texttt{t}}$. Moreover, 
let $\mathbf{f}_{\texttt{r}}$ be the registration field that maps coordinates of $I_{\texttt{s}}$ to
coordinates of $I_{\texttt{t}}$. Given these definitions, the optimization problem can be written as, $\arg \min_{\mathbf{f}_{\texttt{r}}}\int_{\mathcal{X}} \|I_{s}(\mathbf{f}_{\texttt{r}}(x))-I_{t}(x)\|^2dx$.
Often, $\mathbf{f}_{\texttt{r}}$ is characterized by a displacement vector field $\mathbf{u}_{\texttt{r}}$. This vector specifies the vector offset for each
voxel: $\mathbf{f}_{\texttt{r}}=\mathbf{I}_d+\mathbf{u}_{\texttt{r}}$, where $\mathbf{I}_d$ is the identity transform. Hence, the problem transforms to,

\begin{align}
    \min f(\mathbf{u}_{\texttt{r}})= \int_{\mathcal{X}} \|I_{s}(x+\mathbf{u}_{\texttt{r}}(x))-I_{t}(x)\|^2dx\label{eq:registration}
\end{align}

Problem \eqref{eq:registration} is highly non-convex and in many applications, e.g., medical image analysis,
the displacement $\mathbf{u}_{\texttt{r}}$ must satisfy some regularization/smoothness conditions. The extensive prior works have devised various penalty terms to enforce the smoothness of the displacement fields. One such approach is to penalize the gradients' norms of the displacement along the x and y axis. In this paper, we enforce upper and
lower bounds for gradients of $\mathbf{u}_{\texttt{r}}$, thus, the feasible range of
problem \eqref{eq:registration} can be defined as,
\begin{align}
    \mathcal{S}=\{-\epsilon \mathbf{I}_d\leq\mathbf{A}_{\texttt{r}}\mathbf{u}_{\texttt{r}}\leq \epsilon \mathbf{I}_d\}
\end{align}
where $\mathbf{A}_{\texttt{r}}$ denotes the gradient operator along x\&y axis. We choose $\epsilon=0.01$ in the paper.

\begin{figure}[htbp]
\centering
\setlength{\abovecaptionskip}{0.cm}
\includegraphics[width=0.98\columnwidth]{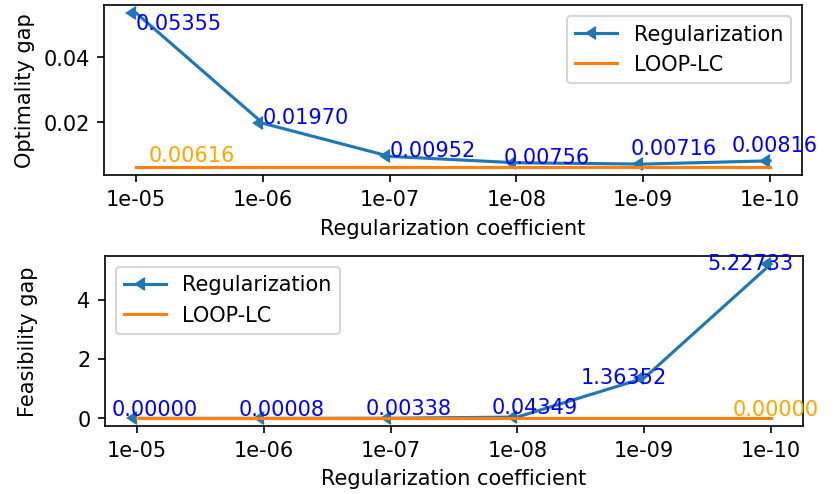}
\caption{Optimality and feasibility results of image registration problems. Optimality gap is measured as $\frac{1}{N}
    \sum_{i=1}^{N} f(\mathbf{u}^{(i)}_{\texttt{r}})$, while the feasibility gap is calculated as $\frac{1}{N}$ $\left ( \left \| \max (\mathbf{A}_{\texttt{r}}\mathbf{u}_{\texttt{r}}- \epsilon \mathbf{I}_d,\mathbf{0})\right \|_1 +\right.$ $ \left.  \left \| \max (-\mathbf{A}_{\texttt{r}}\mathbf{u}_{\texttt{r}}- \epsilon \mathbf{I}_d,\mathbf{0})\right \|_1 \right )$.}
\label{f:image_of}
\end{figure} 

\textbf{Dataset}: We use 25000 pairs of images from the MNIST dataset \cite{lecun1998gradient} for training. 

\textbf{Comparison}: We compare \LOOPLC~against the regularization method \cite{mao2017least} that utilizes mean squared error as the penalty term.

\textbf{Parameters}: We use a customized ResNet  \cite{he2016deep} as the neural network architecture. The Tanh activation is added to the output layer for \LOOPLC~model. The learning rate is set to 0.001.

\textbf{Interior point finder}:  $\mathcal{S}^{\texttt{Ref}}$ is fixed in the image registration problem. Therefore, we use $\mathbf{0}$ as the interior point in \LOOPLC.

\begin{figure}[htbp]
\centering
\vspace{-.4cm}
\setlength{\abovecaptionskip}{0.cm}
\includegraphics[width=0.98\columnwidth]{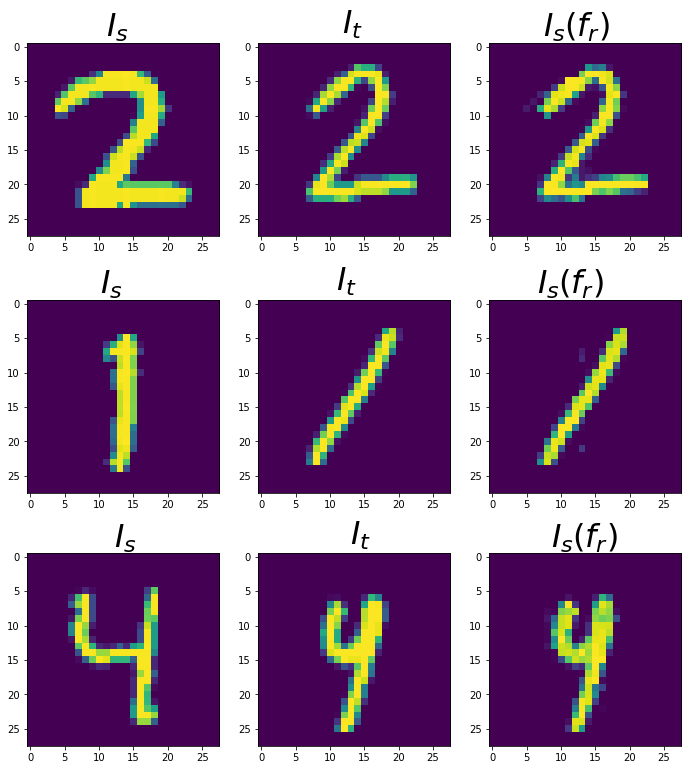}
\caption{Examples of learning results of image registration problems. The source image and target image are in
columns 1-2, and the results 
using \LOOPLC~model in column 3. A well-tuned registration function will produce $I_{s}(\mathbf{f}_{\texttt{r}})$(column 3) similar to images in column 2. Our models perform well over various
images while maintaining smooth displacements.
}
\label{f:r}
\end{figure}

\textbf{Results}: The average per-instance time (using GPU) for regularization method is $1.5576*10^{-5}$ seconds while  \LOOPLC~outputs the results in $1.7699*10^{-5}$ seconds. Optimality and feasibility results are shown in Figure \ref{f:image_of}. Although the regularization method slightly outperforms the \LOOPLC~in terms of speed, it yields a considerable feasibility gap. 
The \LOOPLC~model, which is free of parameter tuning, achieves high-quality (close-to-optimal) solutions while guaranteeing feasibility with respect to hard constraints. Figure \ref{f:r} illustrates training results using \LOOPLC~model.

\vspace{-.7cm}
\section{Conclusion}
This paper introduces the \LOOPLC~ model for solving an optimization problem with hard
linear constraints. At its core, our method is a neural approximator
that maps the inputs to an optimization problem with hard
linear constraints to a high-quality feasible solution (near optimal). In a nutshell, our proposed model learns a neural approximator that maps the optimization’s inputs to an optimal point in the $\ell_\infty$-norm unit ball and then maps the
$\ell_\infty$-norm unit ball to the feasible set of the original problem through a gauge map. Unlike current learning-assisted solutions, our
method is free of parameter-tuning and removes iterations altogether. Our results on convex and non-convex optimization tasks showcase that the \LOOPLC~ achieves close-to-optimal feasible solutions(with respect to hard constraints) while outperforming existing solutions in terms of solution time. Our proposed method is especially applicable to complex optimization problems with simple linear constraints where the interior points could be quickly produced.

\bibliography{aaai23.bib}

\appendix

\section{Alternative Methods to find interior points}

\subsection{Basic feasible points theory method}

\begin{figure}[htbp]
\centering
\setlength{\abovecaptionskip}{0.2cm}
\includegraphics[width=0.7\columnwidth]{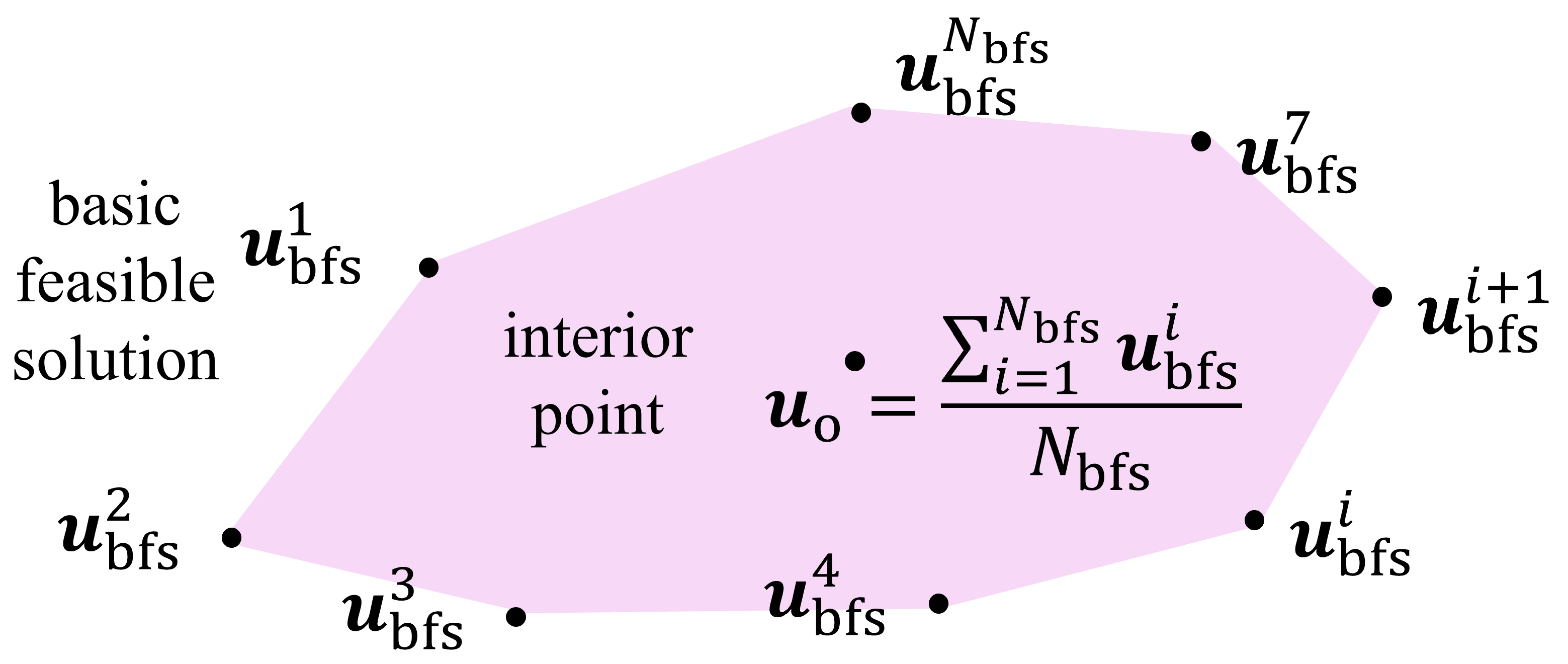}
\caption{The basic feasible points theory method uses the basic feasible points to find interior points, i.e., an interior point can be $\mathbf{u}_{\texttt{o}}=\sum_{i=1}^{8}\mathbf{u}^i_{\texttt{bfs}}/8$, here $\mathbf{u}^i_{\texttt{bfs}}$ represents a basic feasible point.}
\label{f:bfs}
\end{figure}

The basic feasible solution theory  \cite{matousek2006understanding} in linear programming provides an efficient way to find interior points. Geometrically, each basic feasible solution corresponds to a corner of the feasible solutions' polyhedron. According to the features of convex sets, the average of basic feasible solutions is an interior point as long as the algebraic
interior exists \cite{matousek2006understanding} (as shown in Figure \ref{f:bfs}): first, find the 
basic feasible solutions and then use their average as the interior point.

In what follows, we will describe the fundamentals of finding basic feasible points of feasible space $\mathcal{S}^{\texttt{Ref}}$. We first use the slack variables $\mathbf{z}\in \mathbb{R}^{N_I}$ to convert inequality of $\mathcal{S}^{\texttt{Ref}}=\left \{ \mathbf{u}^{\texttt{Indep}}|\mathbf{A}\mathbf{u}^{\texttt{Indep}}+\mathbf{B}\mathbf{x}+\mathbf{b} \right \}\leq0$ into an equational form.

\begin{align}
    \mathbf{A}\mathbf{u}^{\texttt{Indep}}+\mathbf{B}\mathbf{x}+\mathbf{b}+\mathbf{z} =0\nonumber\\
    \mathbf{z}\geq 0\label{eq:equational form}
\end{align}

Then, let's remove $\mathbf{u}^{\texttt{Indep}}$. To this end, we select $(N_{\texttt{opt}}-N_{\texttt{eq}})$ linearly independent rows in $\mathbf{A}$ to form $\mathbf{A}^{\texttt{Indep}}$ and let $\mathbf{I}^{\texttt{Indep}}$ denote the index matrix of the linearly independent rows s.t. $\mathbf{A}^{\texttt{Indep}}=\mathbf{I}^{\texttt{Indep}}\mathbf{A}$. The equality in \eqref{eq:equational form} turns to: $\mathbf{A}^{\texttt{Indep}}\mathbf{u}^{\texttt{Indep}}+\mathbf{I}^{\texttt{Indep}}(\mathbf{B}\mathbf{x}+\mathbf{b})+\mathbf{I}^{\texttt{Indep}}\mathbf{z} =0$. Then, we can derive the relationship between $\mathbf{u}^{\texttt{Indep}}$ and the slack variables $\mathbf{z}$:
\begin{align}
    \mathbf{u}^{\texttt{Indep}}&=\mathbb{H}(\mathbf{z},\mathbf{x})\nonumber\\
    &=-{\mathbf{A}^{\texttt{Indep}}}^{-1}\mathbf{I}^{\texttt{Indep}}(\mathbf{B}\mathbf{x}+\mathbf{b})-{\mathbf{A}^{\texttt{Indep}}}^{-1}\mathbf{I}^{\texttt{Indep}}\mathbf{z} \label{eq:u_z} 
\end{align}

Substitute \eqref{eq:u_z} to \eqref{eq:equational form}, then \eqref{eq:equational form} turns to \eqref{eq:simple form}, where $\mathbf{\hat{A}}=(\mathbf{I}-\mathbf{A}{\mathbf{A}^{\texttt{Indep}}}^{-1}\mathbf{I}^{\texttt{Indep}})$, $\mathbf{\hat{B}}=(\mathbf{I}-\mathbf{A}{\mathbf{A}^{\texttt{Indep}}}^{-1}\mathbf{I}^{\texttt{Indep}})\mathbf{B}$, $\mathbf{\hat{b}}=(\mathbf{I}-\mathbf{A}{\mathbf{A}^{\texttt{Indep}}}^{-1}\mathbf{I}^{\texttt{Indep}})\mathbf{b}$, $\mathbf{I}$ is the identity matrix.
\begin{align}
  \mathbf{\hat{A}}\mathbf{z}+\mathbf{\hat{B}}\mathbf{x}+\mathbf{\hat{b}}=0\nonumber\\
    \mathbf{z}\geq 0\label{eq:simple form}
\end{align}

Assume $\textup{rank}(\mathbf{\hat{A}})=N_{\hat{A}}$. 
Then, we select $N_{\hat{A}}$ linearly independent columns in $\mathbf{\hat{A}}$.
There can be several such combinations($i=1,...,N_{index}$) of linearly independent columns, and we use use $\mathbb{\tilde{I}}^i$ to 
denote the index set of $i$th
combination of linearly independent columns $\mathbf{\hat{A}}^{\texttt{Indep}^i}$. Respectively, other linearly dependent columns form $\mathbf{\hat{A}}^{\texttt{Dep}^i}$. We could also divide $\mathbf{z}$ and $\mathbf{\hat{b}}$ according to index set $\mathbb{\tilde{I}}^i$ as $\begin{bmatrix}
\mathbf{\hat{A}}^{\texttt{Indep}^i} & \mathbf{\hat{A}}^{\texttt{Dep}^i}
\end{bmatrix}
\begin{bmatrix}
\mathbf{z}^{\texttt{Indep}} \\
 \mathbf{z}^{\texttt{Dep}}
\end{bmatrix}=-(\mathbf{\hat{B}}\mathbf{x}+\mathbf{\hat{b}})$.


Let $ \mathbf{z}^{\texttt{Dep}}=\mathbf{0}$. According to Proposition 4.2.2 in  \cite{matousek2006understanding}, we identify a basic feasible point  when $\mathbf{z}^{\texttt{Indep}}=-(\mathbf{\hat{A}}^{{\texttt{Indep}}^{i}})^{-1}(\mathbf{\hat{B}}\mathbf{x}+\mathbf{\hat{b}})\geq \mathbf{0}$. Put differently, the condition of identifying  
the basic feasible point is that the flag variables $\mathbf{z}$
satisfy:

\begin{align}
    \mathbf{z}=\begin{bmatrix}
\mathbf{z}^{\texttt{Indep}} \\
 \mathbf{z}^{\texttt{Dep}}
\end{bmatrix}=\begin{bmatrix}
\mathbf{z}^{\texttt{Indep}} \\
 \mathbf{0}
\end{bmatrix}=\begin{bmatrix}
\mathbf{\tilde{A}}^{i}\mathbf{x}+\mathbf{\tilde{b}}^{i} \\
 \mathbf{0}
\end{bmatrix}\geq \mathbf{0}\label{eq:condition}
\end{align}
where $\mathbf{\tilde{A}}^i=-(\mathbf{\hat{A}}^{{\texttt{Indep}}^{i}})^{-1}\mathbf{\hat{B}}^{\texttt{Indep}^i}$ and $\mathbf{\tilde{b}}^i=-(\mathbf{\hat{A}}^{{\texttt{Indep}}^{i}})^{-1}\mathbf{\hat{b}}^{\texttt{Indep}^i}$.

Therefore, we just need to find all index sets $\mathbb{\tilde{I}}^i$ and compute $\mathbf{z}^{\texttt{Indep}}$ accordingly to predict whether it's a basic feasible point.
Further, we calculate $\mathbf{u}^{\texttt{Indep}}$ according to \eqref{eq:u_z}.
Finally, the average $\mathbf{u}_{\texttt{o}}$ of all the basic feasible points is an interior point according to the feature of convex sets.

Note that $\mathbb{\tilde{I}}=\left \{ \mathbb{\tilde{I}}^i,i=1,...,N_{index} \right \}$ is independent of data $\mathbf{x}$. Therefore only one $\mathbb{\tilde{I}}$ exists given problem \eqref{eq:problem1} and we don't need to compute a new index set given different $\mathbf{x}$.

Implementation of this method is straightforward, but the computation will exponentially increase with more constraints.


\subsection{Two-phase method to learn the interior points}

Considering that problem \eqref{eq:problemif} is also a linearly-constrained problem, \LOOPLC~model can be leveraged to learn the solution. Since  $\left \| \mathbf{A}\mathbf{u}^{\texttt{Indep}}+\mathbf{B}\mathbf{x}+\mathbf{b} \right \|_2\neq 0, \forall \mathbf{x}$, in most practical settings, we have:
\begin{align}
  \mathbf{A}\mathbf{u}^{\texttt{Indep}}+\mathbf{B}\mathbf{x}+\mathbf{b}  \nonumber\\
  \leq \mathbf{1}\left \| \mathbf{A}\mathbf{u}^{\texttt{Indep}}+\mathbf{B}\mathbf{x}+\mathbf{b} \right \|_2\nonumber\\
  < 2*\mathbf{1}\left \| \mathbf{A}\mathbf{u}^{\texttt{Indep}}+\mathbf{B}\mathbf{x}+\mathbf{b} \right \|_2
\end{align}
Therefore, $\forall \mathbf{u}^{\texttt{Indep}}$, 
$\begin{bmatrix}
\mathbf{u}^{\texttt{Indep}}\\ 
\left \| \mathbf{A}\mathbf{u}^{\texttt{Indep}}+\mathbf{B}\mathbf{x}+\mathbf{b} \right \|_2
\end{bmatrix}$ is an interior point for \eqref{eq:problemif}. 

\begin{figure}[htbp]
\centering
\setlength{\abovecaptionskip}{0.cm}
\includegraphics[width=1\columnwidth]{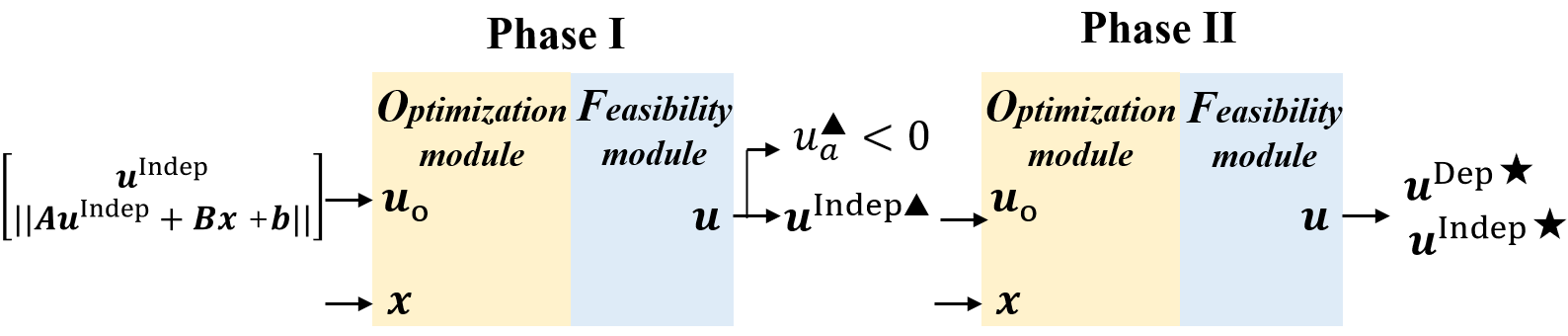}
\caption{Two-phase method to find interior points.}
\label{f:twophase}
\end{figure}

Further,  we propose a two-phase 
the scheme as in Figure \ref{f:twophase}, where \LOOPLC~model is first applied to \eqref{eq:problemif} with $\begin{bmatrix}
\mathbf{u}^{\texttt{Indep}}\\ 
\left \| \mathbf{A}\mathbf{u}^{\texttt{Indep}}+\mathbf{B}\mathbf{x}+\mathbf{b} \right \|_2
\end{bmatrix}$ as the interior point. If the solution $u_a^{\blacktriangle  }<0$, then \LOOPLC~model will be applied again in Phase II to solve problem \eqref{eq:problem1-abst.reform}. In the Phase II, $\mathbf{u}^{\texttt{Indep}\blacktriangle }$ is the interior point for \eqref{eq:problem1-abst.reform}.

The two-phase method will be extremely fast, and the  execution time will be only twice the time of a single forward process using \LOOPLC~model. The two-phase method is also free of any solvers. However, the solution $u_a^{\blacktriangle  }$ might be larger than 0 in Phase I due to prediction error. 

\section{The complexity of the constraint set}
\begin{figure}[htbp]
\centering
\setlength{\abovecaptionskip}{0.cm}
\includegraphics[width=1\columnwidth]{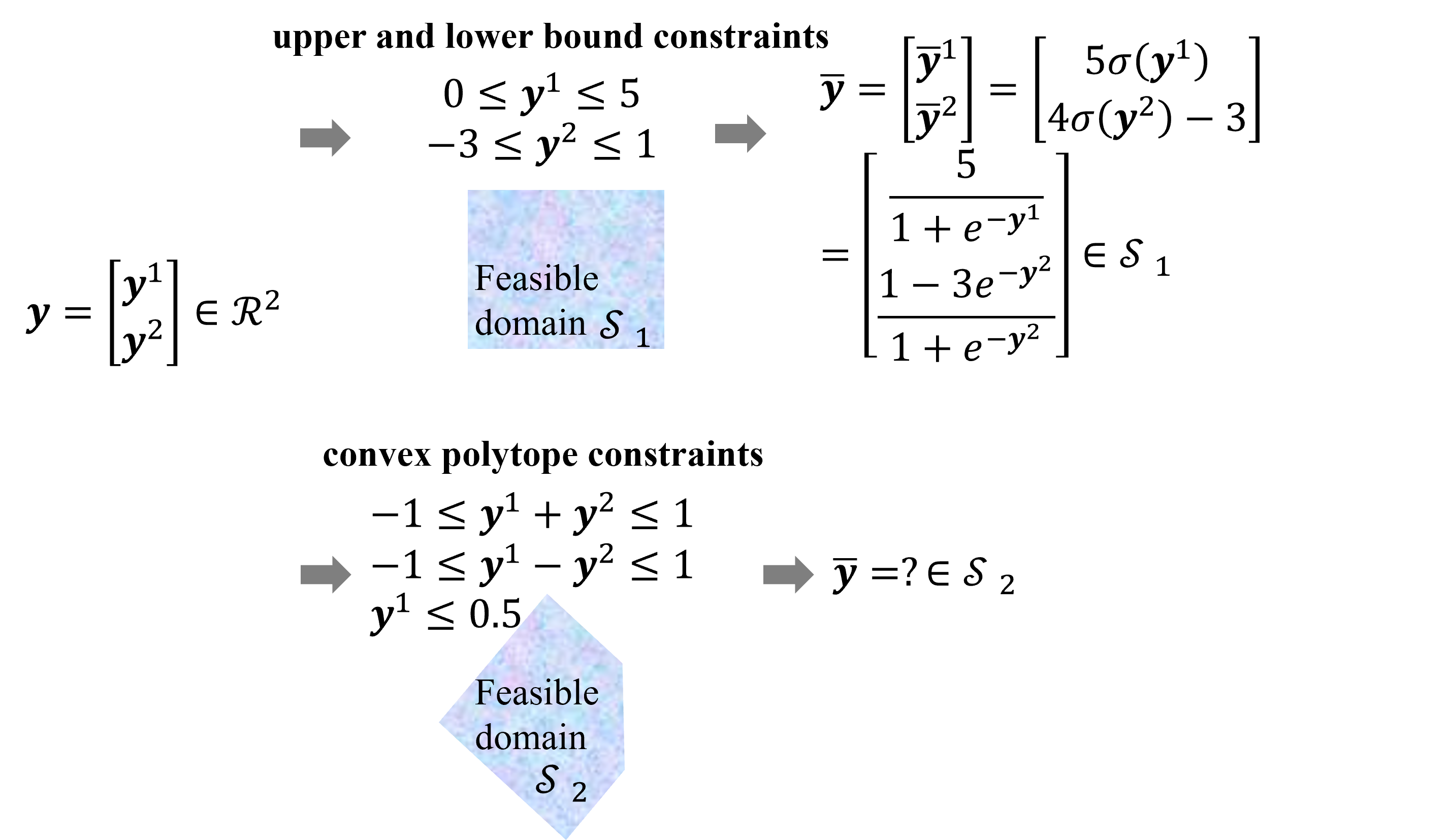}
\caption{This figure provides an example of enforcing upper and lower bounds and coupled linear constraints. Note, applying Logistic activation results in keeping the output variable inside the feasible domain $\mathcal{S}_1$. However, it is challenging to ensure the feasibility in the case of $\mathcal{S}_2$ (that includes coupled constraints and variables).}
\label{f:exureample}
\end{figure}

As mentioned earlier, it is relatively straightforward to guarantee feasibility if the feasible range only constitutes upper and lower bounds. Adopting any activation function that produces a bounded output can enforce optimization variables to stay within pre-determined upper and lower bounds. For example, Figure. \ref{f:exureample} shows that applying Logistic activation could enforce the output variable within the feasible domain $\mathcal{S}_1$.
However, it is challenging to utilize neural approximators to solve \eqref{eq:problem1-abst.} where the feasible range includes coupled constraints and variables, as $\mathcal{S}_2$ in
Figure. \ref{f:exureample}.

\end{document}